\documentclass[apjl]{emulateapj}

\usepackage[dvips]{epsfig}
\usepackage{natbib}
\usepackage{apjfonts}

\begin{document}

\title{Energy Injection Episodes in Gamma Ray Bursts: The Light Curves 
and Polarization Properties of GRB~021004\altaffilmark{1}}

\author{G.~Bj\"ornsson, E.~H.~Gudmundsson, G.~J\'ohannesson\altaffilmark{2}}

\altaffiltext{1}{Some of the radio data referred to in this paper were
  drawn from the GRB Large Program at the VLA, {\tt
    http://www.vla.nrao.edu/astro/prop/largeprop/.} 
NRAO is a facility of the National Science
  Foundation operated under cooperative agreement by Associated
  Universities, Inc.} 
\altaffiltext{2}{Science Institute, University
  of Iceland, Dunhaga~3, IS--107 Reykjavik, Iceland, e-mail:gulli,
  einar, gudlaugu@raunvis.hi.is}

\begin{abstract}
  Several GRB afterglow light curves deviate
  strongly from the power law decay observed in most bursts. We show
  that these variations can be accounted for by including refreshed
  shocks in the standard fireball model previously used to interpret
  the overall afterglow behavior. As an example we
  consider GRB~021004 that exhibited strong light curve variations and
  has a reasonably well time-resolved polarimetry. We show that the
  light curves in the $R$-band, X-rays and in the radio can be
  accounted for by four energy injection episodes in addition to the
  initial event. The polarization variations are shown to be a
  consequence of the injections.
\end{abstract}


\keywords{gamma rays: bursts --- gamma rays: theory --- polarization}

\section{Introduction}
\label{sec:intro}

A long-duration gamma-ray burst (GRB) is now generally believed to
occur following the core collapse of a massive star
\citep{Woosley1993, Stanek2003, Hjorth2003}. Following the collapse,
the released energy pierces a hole through the star along its rotation
axis, sweeps up and shocks the surrounding interstellar medium 
and produces a long-lived afterglow emission. According to the
standard fireball model \citep{Meszaros2002, ZhangMesz2003,
  Piran2004}, the expanding fireball slows down as it sweeps up more
and more ambient material and decays in brightness.

The standard fireball model (SFM) predicts that the afterglow light
curves should be power laws in time, with a break due either to one of
the characteristic frequencies of synchrotron radiation moving through
the observing band \citep[e.g.][]{SarPirNar1998} or because of the
outflow being collimated \citep{Rhoads1999}. The former would only
give rise to a modest steepening, while in the latter case it can be
substantial in addition to being achromatic. In most cases where a
break has been observed, the steepening is substantial and achromatic
in the optical domain and the latter explanation appears to be the
appropriate one.

For the first GRB afterglow light curves observed, the SFM
provided adequate interpretation of the data. In recent years, several
well observed afterglows have shown strong deviation from smooth
power-laws, e.g.\ GRB~011211 \citep{Jakobsson2004}, GRB~021004
\citep{Fox2003, Holland2003}, and GRB~030329 \citep{Price2003,
  Matheson2003b, Lipkin2004}. Bursts like GRB~000301C, GRB~020813 and
GRB~010222 also deviate from the standard behavior as their initial
light curve decay rate was slower than can be easily accommodated
within the model. In these cases, a continuous energy injection may
explain the slow decay as has been suggested for GRB~010222
\citep{Bjornsson2002}.

For light curves showing 'bumps' and 'wiggles' during the first few
days, as e.g.\ GRB~021004, it has been suggested that these might
arise when the fireball encounters density irregularities in
the ambient medium \citep{Lazzati2002, HeylPerna2003, NakarPiran2003}.
Higher density regions would cause brightening episodes, while a
rarification immediately following the higher density appears
necessary to get the light curve decay 'back on track'. 
\cite{Fox2003} suggested that the light curve variability
of GRB~021004 may be due to refreshed shocks. Recently, it has been
claimed that such models can be rejected as they are unable
to explain simultaneously the bumpy light curves and the polarization
as the latter would not be affected by the freshly injected energy
\citep[][hereafter L03]{Lazzati2003}.

In this Letter we show, contrary to the conclusions of
L03, that several energy injection episodes may in fact
explain the afterglow light curve re-brightenings
of bursts like GRB~021004, each injection contributing to the light
curve as it catches up with the previously expanding shock front
\citep[see also][for a similar conclusion on GRB~030329]{GranNakPir2003}.  
We consider injections in the
SFM and apply our model to GRB~021004. We
show that not only are the light curves readily accounted for, but
also that its polarization properties follow directly from the model.
Detailed account of this work as well as application to other bursts
will be presented elsewhere (J\'ohannesson et~al.\ in preparation).

\section{Energy Injection}
\label{sec:injection}

For a general review of the SFM we refer the
reader to \cite{Meszaros2002}, \cite{ZhangMesz2003}
and \cite{Piran2004}.  The basic idea behind the model is the
self-similar relativistic shock solution of \cite{BM1976}, where it is
assumed that energy is released instantaneously at the onset of the
burst. It is, however, both natural and expected that the energy
release may be either continuous or episodic
\citep{ReesMesz1998,KumarPiran2000,SariMesz2000}.

We extend the SFM as suggested by \cite{ReesMesz1998} by
applying energy and momentum conservation, as in \cite{Rhoads1999}, to
both discrete and continuous energy injections although only discrete
injections will be considered here.  We assume that several discrete
shells are injected simultaneously with different Lorentz factors. The
shell with the highest Lorentz factor, $\Gamma_0$, drives the initial
evolution of the afterglow.  Once it has decelerated to a value lower
than the Lorentz factor of the next shell, they will collide with a
delay corresponding to the time it takes the second shell to reach the
first.  The collision is assumed to be instantaneous and its dynamics
is neglected as is any anisotropy that may results from the shell
interactions. In afterglow interpretations, the delay is fixed by the
start of a brightening episode. The energy of the shell is determined
by the observed increase in flux level.  Similar results can be
achieved with a continuous energy injection as discussed by
\cite{PanMeszRees1998}, but requires a very steep energy injection
profile and in some cases even a varying electron energy index, $p$.

In calculating light curves and spectra, we assume that the
radiation is of synchrotron origin, and we consider the local synchrotron
spectra at each point in the outflow to consist of power law segments,
smoothly joined at the characteristic frequencies \citep{GranSar2002}.
We integrate over a thin shell at the equal arrival time surface.
Each shell element is assumed to be locally homogeneous, its
thickness being determined by the jump conditions across the shock
\citep{BM1976} and the conservation of particles.
We can therefore consider general density profiles, for example a
constant density environment, a wind or various density
irregularities. 

We calculate the instantaneous fireball polarization as in 
\cite{Rossi2004}. We assume a random magnetic field compressed
by the blast-wave, thereby introducing some alignment perpendicular to
the compressed direction \citep{Laing1980}. We evaluate at a given
observer time the contribution of each surface element of the equal
arrival time surface to the Stokes parameters
$dU=P(\theta)dL\cos(2\phi)$, and $dQ=P(\theta)dL\sin(2\phi),$ with the
angular dependence of the polarization given by
$P(\theta)=P_0\sin^2\theta'/(1+\cos^2\theta')$ \citep{Laing1980}.
Here, $\theta'$ is the angle from the line of sight in the comoving
frame and $dL$ is the local luminosity. In the optical range for most
of the fireball evolution, the maximum degree of polarization is taken
to be $P_0=(p+1)/(p+7/3)\approx 70\%$ for $p=2.2$ \citep[see][for
details]{RybLight1979}.  For a measurable polarization to occur, 
the line of sight to the observer has to be off the outflow
symmetry axis \citep{GL1999, Sari1999}.  As in \cite{GL1999}, we find
that for an instantaneous energy release, the polarization light curve
has two extrema, bracketing the time when $\theta\approx1/\Gamma$. At
approximately that time the polarization angle rotates by 90$^\circ$
\citep{GL1999,Sari1999,GranotKonigl2003}.  It is important to note
that the maximum degree of polarization and time of change of the
polarization angle depend strongly on the viewing angle. The
polarization evolution depends mostly on the evolution of $\Gamma$,
but is also affected by the evolution of the jet opening angle.
\cite{Rossi2004} have shown that the larger the lateral velocity, the
lower the polarization. We assume the sideways expansion is given
by the comoving sound speed, $c_s=c/\sqrt{3}$.  Recent work has shown
that the lateral expansion may in fact be slower \citep{Granot2001},
but we adopt this assumption to highlight the effects of the energy
injections.

The number of model parameters can be quite large if there are many
brightenings. The global parameters: initial energy, $E_0$, Lorentz
factor, $\Gamma_0$, half-opening angle, $\theta_0$, ambient density,
$n_0$ (or density profile), relative energy density in electrons,
$\epsilon_e$, and magnetic field, $\epsilon_B$, and electron index,
$p$, are determined, as in the SFM, from the initial afterglow
evolution and the total flux level (using the burst redshift).  All
episode parameters are fixed by the observed time of brightenings and
the increase in flux levels. As discussed by \cite{ZhangMesz2002}
(hereafter ZM02), only the energy ratios and the relative velocities
are relevant as long as the relativistic phase lasts, and therefore
absolute shell energies and Lorentz factors are not needed.  If
polarimetry is available this may be the most reliable way to
determine the so called jet break time as the polarization angle is
predicted to change by $90^\circ$ at that time.  Determining the break
time from a broken power-law fit to the optical light curve can lead
to an erroneous result if the light curve is not smooth, as e.g.\ in
the case of GRB~021004. In addition, for the interpretation to be
consistent, the polarization levels predicted by the model should
agree with those observed, once all parameters have been determined.
Our code has been extensively tested and compared to other work.  We
have also compared the output with analytical results, and find good
agreement in parameter ranges where the assumptions used in analytical
work are valid.

The effects of a single injection episode can be described as follows:
When a relativistic shell
catches up with the slower shock front propagating into the ambient
medium, it increases the $\Gamma$ of the forward shock (ZM02), 
but the subsequent evolution of $\Gamma$ continues
with the same decay rate as before.
Reverse shocks may be expected,
but these are expected to be weak for mild energy injections
(ZM02).
The flux from the fireball increases, but as the energy addition is
instantaneous the light curve decay will from then on also continue
with the same rate as prior to the shell injection.
The rise in flux
will not be sharply defined in time as the flux is obtained by
integration over the equal arrival time surface that smooths out the
transition.  The injected shell will thus result in a smooth bump in
the light curve.  The net result is that the light curve is 'shifted'
upwards at the time of the injection, but retains it original decay
rate behavior from then on, the rate being most strongly determined by
$p$, and the density structure of the ambient medium
\citep[e.g.][]{PanMeszRees1998}.
Each energy injection episode delays the decaying evolution of
$\Gamma$, and causes it to increase temporarily. In addition, the
emitting region of the relativistic outflow centered on the line of
sight, temporarily brightens up and outshines the bright emitting
ring-like region around the line of sight \citep[see][for a detailed
discussion]{Waxman1997, PanMesz1998, GranPirSar1999}. As a result the
flux will increase slowing the light curve decay
\citep{KumarPiran2000}.  It is important to realize that contrary to
earlier statements (L03), polarization will
also be affected by the energy injection, as increasing $\Gamma$
increases the flux, causes increased aberration, decreases the
emitting surface area and thus decreases the degree of polarization
compared to an unperturbed evolution. We will show examples of these
effects in the next section.  Additional injections can be viewed as
superpositions of repeated single episodes with similar effects on the
light curve.


\begin{table}[t]
\begin{center}
\caption{Model parameters for GRB~021004}
\begin{tabular}{lr}
Parameter       &    Value      \\
\hline
$E_0$           &  $1.0$      \\
$\Gamma_0$      &  $800$      \\
$p$             &  $2.2$      \\
$n_0$           &  $26.0$       \\
$E_1$           &  $3.5$         \\
$E_2$           &  $5.6$         \\
$E_3$           &  $13.0$         \\
$E_4$           &  $7.0$         \\

\hline
\label{table1}
\end{tabular}
\end{center}
Initial half-opening angle is $\theta_0=1.4^\circ$,
the line of sight angle is $\theta_v=0.95\theta_0$,
$\epsilon_e=0.21$, and $\epsilon_B=2\cdot 10^{-4}$. 
$n_0$ is in units of cm$^{-3}$, $E_0$ is in units 
of $10^{50}$ ergs, and the four energy injection values, 
$E_i$, are relative to $E_0$. The four injection times 
in the observer frame are approximately 
1\,h, 16\,h, 42\,h and 105\,h. 
\end{table}


\section{GRB~021004}

As an example, we consider GRB~021004. It showed strong light curve
variations with a best fit light curve break time of $4.74$ days
\citep{Holland2003}. This break time was obtained by fitting a broken
power law to the data and resulted in a rather large formal error.
The afterglow also exhibited variable polarization levels and a
90$^\circ$ change in polarization angle at approximately $1.0$ day
\citep{Rol2003}. These two time estimates, if interpreted
within the SFM, should be similar but are in this case
inconsistent with each other.


\begin{figure}[t]
\plotone{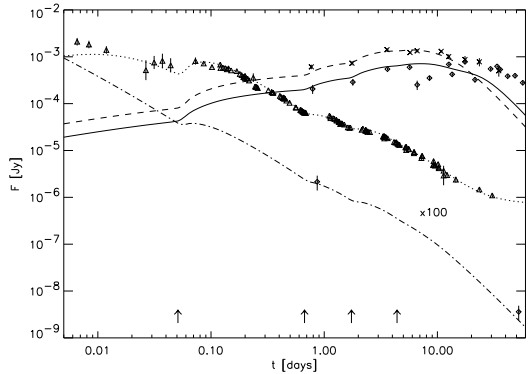}
\caption{Light curves of GRB~021004: $R$-band (dotted), 8.46 GHz (solid), 
  22.5 GHz (dashed), and $X$-ray (multiplied by a factor of 100;
  dash-dotted). Variability in the optical light curve is obtained
  by imposing 4 energy injection episodes at times indicated by the
  arrows.  Parameters were adjusted so that the model light curves
  would also go through the $X$-ray and radio light curves as well as
  the polarization data. A host galaxy of $R$-band magnitude 24 was
  added to the model optical light curve.  $R$-band data is from
  \cite{Uemura2003, Pandey2003, Holland2003}.  X-ray data is from
  \cite{SakoHarrison2002a, SakoHarrison2002b}.  Radio data is from
  \cite{Berger02}; \cite{Frail2002} and the GRB Large Program at the
  VLA (AK509)$^1$. Radio data hints at a 5th injection at 
  11-12 days.}
\label{fig:fig1}
\end{figure}


We find that the light curves can be explained by 4 injection
episodes, superimposed on the initial GRB event. We summarize the
model parameters in Table~\ref{table1}, and show the model light
curves in Fig.~\ref{fig:fig1}. We set $p=2.2$, as theoretical studies
suggest this to be a universal value \citep[see][for a review and
references]{Piran2004}.  Using the burst redshift, $z=2.335$
\citep[e.g.][]{Moller2002}, $E_0$, $\Gamma_0$ and $n_0$, are
determined by demanding that the model flux matches that observed in
the same way as in the standard model. The bumps in the $R$-band light
curve are then used to set the times and relative energies of the
injection episodes.  Other parameters, such as $\epsilon_e$ and
$\epsilon_B$, are then adjusted until the model agrees with the radio
and $X$-ray data.  Finally, the change in polarization angle to fixes
the opening angle of the jet, $1.4^\circ$. Recall that this depends on
the rate of lateral expansion, here assumed to be constant.

All four injections are mild (ZM02), the relative
Lorentz factor in all cases about 2.  The relative energy of the first
injection is 3.5, in subsequent episodes it is 1.24, 1.27 and about
0.3 of the total in the fourth episode.  The peak flux from the
reverse shocks is maximum for the first event, a factor of 30 higher
than from the forward shock with the frequency at maximum reverse flux
of about 2.5 GHz. No radio data is available at the time of the
first injection.  All subsequent reverse shocks have lower maximum
fluxes and frequencies in the sub-GHz range. 
We therefore neglected the reverse shock contributions to the flux.

As seen in Fig.~\ref{fig:fig1}, the model is able to account for the
light curves in all observed wavelength regions at all times,
except for the first half hour or so when an initial reverse shock may
be dominating the flux \citep{KobayashiZhang2003,
  PanaitescuKumar2004}, and the radio data after about 20 days. A
fifth injection may be able to account for the late radio data
followed by a transition to a non-relativistic expansion regime at 50-60 days.


\begin{figure}[t]
\plotone{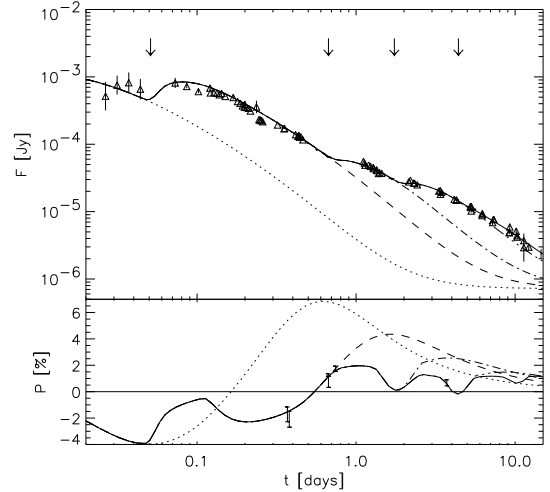}
\caption{A segment of the $R$-band light curve of GRB~021004 (top panel).
  Effects of each of the energy injection
  events is shown. Dotted curve shows the theoretical afterglow
  model without additional energy injection. Dashed curve shows
  the effect of one injection event, dot-dashed two events,
  dash-triple-dotted three events and the solid curve shows the effect of
  all four injection events.  Arrows indicate the injection times.
  Lower panel shows the corresponding polarization light curves.
  Note how each injection episode causes the polarization to deviate
  from the curve of unperturbed evolution. Polarimetry data is
  from \cite{Rol2003} and L03.}
\label{fig:fig2}
\end{figure}

The intrinsic optical and X-ray spectral slopes are fixed by our
choice of $p=2.2$. At 1.4 days these are $\beta_{\rm opt}=(p-1)/2=0.6$
and $\beta_{\rm X}=p/2=1.1$.  The former is within 2$\sigma$ of the
intrinsic (extinction corrected) spectral index, estimated in
\cite{Holland2003}, of $\beta_{UH}=0.39\pm0.12$.  \cite{Matheson2003a}
find a steeper slope of the optical spectra but with a clear
curvature, $\beta=0.98\pm0.03$.  \cite{Holland2003} also obtained an
excellent fit to the 2--10 keV {\em Chandra} spectral index of
$1.03\pm0.06$, agreeing with the model prediction. In the 0.4--10 keV
range and including absorption, they find $\beta_{\rm X}=0.94\pm
0.03$.  These results are consistent with those of \cite{Butler2003}
and \cite{SakoHarrison2002a}.

In Fig.~\ref{fig:fig2} we plot the $R$-band and polarization light
curves of GRB~021004. We show the effect of each injection episode
separately as well as the combined result. With this interpretation,
the estimate by \cite{Holland2003} of the late jet break time now has
a simple explanation. The repeated energy injections slow the early
light curve decay and delay the steepening until after the last
injection. A broken power law thus underestimates the pre-break decay
slope and overestimates the jet break time.  The change in
polarization angle should give a more reliable estimate of the break
time.

As mentioned above, with only one energy release event, 
we recover the polarization behavior of \cite{GL1999}
with differences due to our more detailed fireball model (dotted
curves).  Adding one injection episode clearly shows the effects on
the optical and polarization light curves (dashed in
fig.~\ref{fig:fig2}). At the time of injection the polarization level
drops sharply and reaches a minimum just after maximum brightness in
the light curve. In this particular burst, the first injection occurs
before the jet break, and because the injection modifies the
temporal evolution of $\Gamma$, it delays the jet break time compared
to a single event, and hence also the 90$^\circ$ change
in polarization angle. The break time, as defined by the change in
polarization angle, is approximately at 0.6 days. This is just before
the 2nd injection and therefore goes unnoticed in the light curves
until after the last injection.

Including all four injections results in the polarization
light curve shown in the lower panel in
fig.~\ref{fig:fig2} (solid).  There, the effect of each injection episode
is very clear. Depending on the strength of the injections,
the polarization angle may change again, although at a very low
polarization level (see local minimum at approximately 4 days). The
correlation between the observed flux and polarization variations is
in this interpretation seen to be directly rooted in the dynamics of
the outflow.

\section{Discussion}
\label{sec:disc}

We have also applied our code to a SFM modified with density
variations in a homogeneous medium, but without energy injections.  We
modeled the variations with Gaussian profiles as in L03.  The
calculated flux is very sensitive to the number of radiating electrons
and care must be taken in counting them. Only when we specifically
introduced the shock profile of \cite{BM1976}, did we manage to get
sufficient brightenings.  A proper treatment of the effects of density
variations, valid at all observer times and radial ranges of interest
here, requires a numerical solution of the dynamical equations and a
self consistent integration over the emitting region.

\cite{NakarOren2004} have interpreted the GRB~021004 optical data
using the ``patchy shell'' model where the angular distribution of
energy in the shock is inhomogeneous due to random fluctuations.  Its
application requires the angular energy distribution to be specified,
but with suitable parameters the model is able to account for the 
general trends in the data.

A contribution to the polarization may originate in
the Galaxy or within the host. L03 conclude
that for GRB~021004, the host contribution can be accounted for by
using spectropolarimetric data, while the Galaxy may dominate at low
polarization levels. However, the time variation of the polarization
can only originate in the source.

It is natural to assume that the central source releases energy in
several discrete events, essentially simultaneously. The total energy
injected into the collimated outflow inferred from our model is about
$3\cdot 10^{51}$ erg. It is of the same order as estimated in other
bursts after beaming correction \citep{Frail2001}.
Applying beaming correction to the isotropic $\gamma$-ray energy
estimated in \cite{Fox2003}, we find that $1.5\cdot 10^{49}$ ergs were
emitted in $\gamma$-rays. We remind the reader that the narrow opening
angle inferred from our modeling is a consequence of assuming
a constant lateral expansion velocity.  In this interpretation,
most of the electromagnetic energy output in GRB~021004 was emitted at
longer wavelengths.

In addition to GRB~021004, we have considered other burst with highly
variable light curves, including GRB~970508 and GRB~030329. Most cases
considered so far can be accounted for by discrete energy injections,
with GRB~030329 a possible exception. This work will be discussed
separately.

The self similar solution is based on the assumption of a uniform or a
wind structured ambient medium. It may therefore be able to account
for the overall behavior of the expanding shock, but one cannot expect
it to be able to follow density variations in detail.  For that a
numerical solution of the dynamical equations is required.  We
conclude that refreshed shocks {\em can} in fact account for the
observed variability both in the light curves and in the polarization
properties of bursts.
The strongest argument in favor of our interpretation is the fact that 
we are able to account for both broad band behavior as well as the 
polarimetry within a single model.

\acknowledgements
We thank the referee for a constructive report and
P.\ Jakobs\-son for helpful comments on the manuscript.
This work was supported in part by the University of Iceland Research Fund and
a special grant from the Icelandic Research Council.

\end{document}